\documentclass[11pt]{article}

\usepackage{amsmath,amssymb}
\usepackage{graphicx}

\makeatletter
 
  \@addtoreset{equation}{section}
 \makeatother

\newcommand{\be}{\begin{equation}}
\newcommand{\ee}{\end{equation}}
\newcommand{\bea}{\begin{eqnarray}}
\newcommand{\eea}{\end{eqnarray}}
\newcommand{\beann}{\begin{eqnarray*}}
\newcommand{\eeann}{\end{eqnarray*}}
\newcommand{\nn}{\nonumber}
\newcommand{\ba}{\begin{array}}
\newcommand{\ea}{\end{array}}

\newcommand{\e}{\epsilon}

\newcommand{\del}{\partial}

\begin{document}

\renewcommand{\thefootnote}{\arabic{footnote}}
\begin{titlepage}

\begin{center}

\hfill 
\vskip 1.1in

{\bf \Large 
A No-Go Theorem for 
M5-brane Theory} 

\vskip .7in
{\sc 
Chien-Ho CHEN$\,{}^{}$\footnote{chenchienho@gmail.com}, 
Pei-Ming HO$\,{}^{}$\footnote{pmho@phys.ntu.edu.tw} and\ 
Tomohisa TAKIMI$^{}$\footnote{tomo.takimi@gmail.com}}\\
\vskip 10mm
{\it\large
Department of Physics, 
Center for Theoretical Sciences \\
and Leung Center for Cosmology and Particle Astrophysics, \\
National Taiwan University, Taipei 10617, Taiwan,
R.O.C.\\}
\noindent{ \smallskip }\\

\vspace{10pt}
\end{center}
\begin{abstract}

The BLG model for multiple M2-branes motivates an M5-brane theory 
with a novel gauge symmetry defined by the Nambu-Poisson structure. 
This Nambu-Poisson gauge symmetry for an M5-brane in large $C$-field background 
can be matched, 
on double dimension reduction, 
with the Poisson limit of the noncommutative gauge symmetry 
for a D4-brane in $B$-field background. 
Naively, one expects that there should exist a certain 
deformation of the Nambu-Poisson structure 
to match with the full noncommutative gauge symmetry
including higher order terms.
However, 
We prove the no-go theorem that
there is no way to deform the Nambu-Poisson gauge symmetry,
even without assuming the existence of a deformation of Nambu-Poisson bracket,
to match with the noncommutative gauge symmetry in 4+1 dimensions to all order, 
regardless of how the double dimension reduction is implemented.

\end{abstract}

\end{titlepage}

\setcounter{footnote}{0}


\section{Introduction}

Recently, significant progress has been achieved in our understanding of 
multiple M2-branes through the BLG model \cite{BL,Gustavsson} and the ABJM model \cite{ABJM}.
However, the physics of a system of multiple M5-branes still remains 
mysterious, 
although the Lagrangian formulation for a single M5-brane has been known 
for over ten years~\cite{oldM5}.

The description of multiple M5-branes is expected to be a non-Abelian 
generalization of the single M5-brane theory, 
for reasons analogous to why the low energy effective theory for $N$ D-branes 
is a $U(N)$ gauge theory. 
Yet a peculiar property of the multiple M5-brane system is that 
its entropy scales as $N^3$, instead of $N^2$, 
for $N$ coincident M5-branes \cite{n3} (for large $N$).
This suggests that the non-Abelian generalization for M5-branes 
is not an ordinary Yang-Mills theory with the gauge symmetry of 
a certain semi-simple Lie algebra, 
for which the dimension of the Lie algebra scales as $N^2$ 
if the Cartan subalgebra is $N$ dimensional.
Thus the first mystery about M5-branes is: 
{\em What is the gauge symmetry for a system of M5-branes?} 
A no-go theorem was proven in \cite{BHS}, 
saying that a non-Abelian generalization is impossible. 
They considered general deformations of the theory of $N$ commuting copies 
of self-dual 2-form gauge theory in 6 dimensions, 
and found that a consistent non-Abelian deformation is always trivial, 
i.e., equivalent to a change of variables. 
They did not assume the deformed gauge transformations to 
correspond to a particular type of symmetry algebra, 
but only that the deformation is local.

Recently, a new Lagrangian formulation of a single M5-brane 
in a large $C$-field background was introduced \cite{Ho:2008nn}\cite{Ho:2008ve}.
\footnote{A review on the NP M5-brane is given in \cite{Ho:2009zt}, 
and further investigations on this model include
\cite{Bandos:2008fr,Furuuchi:2008ki,Furuuchi:2009zx,Pasti,Furuuchi:2010sp,
Chu:2009iv,Chen:2010jg}.
}
This theory was derived from the BLG model of multiple membranes 
for which the Lie 3-algebra structure was chosen to be the Nambu-Poisson (NP) structure. 
The gauge symmetry is a non-Abelian generalization of 
the Abelian gauge symmetry for a self-dual 2-form gauge potential, 
with the non-Abelian structure introduced through the Nambu-Poisson bracket. 
We will refer to this theory in the following as the {\em NP M5-brane theory}.
In contrast to the old formulation of M5-brane theory, 
the usual Abelian gauge symmetry is promoted to 
the non-Abelian gauge symmetry of the volume-preserving diffeomorphisms (VPD)
in the NP M5 action. 
The coupling constant for the non-Abelian coupling (the NP bracket) 
is inversely proportional to the $C$-field background. 
There is no index going from 1 to $N$, 
and thus the NP M5-brane theory is not 
a counter-example of the no-go theorem \cite{BHS} mentioned above. 
But it reminds us that no-go theorems can always be circumvented. 

Incidentally, the entropy of the system of $N$ M2-branes 
also exhibit a peculiar scaling behavior as $S \sim N^{3/2}$. 
It turns out that, at the level of equations of motion 
(in the absence of an invariant inner product), 
the BLG model can reproduce this scaling behavior using 
a truncated Nambu-Poisson algebra as the Lie 3-algebra \cite{Chu:2008qv}.
We take this as a hint suggesting that further understanding of the NP M5-brane model 
may help us to understand this subject better.

The major evidence that the NP M5-brane action corresponds to 
a large $C$-field background is that it reduces to 
the D4-brane theory in large $B$-field background 
upon double dimension reduction \cite{Ho:2008ve}.
A D4-brane in $B$-field background is known to be described by 
the noncommutative (NC) Yang-Mills action~\cite{SW,Aoki:1999vr}. 
For large $B$-field background, 
the noncommutativity parameter is inversely proportional to 
the $B$-field, i.e., $\theta \simeq 1/B$, 
and the Moyal bracket is approximated by 
the Poisson bracket at the lowest order in the $\theta$-expansion
\be
[ f(x), g(x) ]_{\ast} \equiv f\ast g - g\ast f 
= \theta^{ij} \del_i f \del_j g + {\cal O}(\theta^3). 
\ee
The double dimension reduction of the NP M5-brane 
captures the Poisson bracket structure of the NC D4-brane action, 
but misses all the higher order terms in the $\theta$-expansion.
This means that the NP M5-brane is just a lowest order approximation 
for large $C$-field background in the $1/C$-expansion.
The natural question is thus: 
{\em Can we deform the NP M5-brane theory 
such that its double dimension reduction gives the NC Yang-Mills theory 
to all orders in $\theta$?}

The aim of this paper is to ask a more fundamental question instead:
\begin{quote}
{\em
Can we deform the NP gauge symmetry such that 
it reduces to the NC gauge symmetry 
(including higher order terms) 
upon double dimension reduction?
}
\end{quote}
Obviously, if the answer to this question is negative, 
the answer to the former question must be, too.
Notice that this latter question can be asked without referring 
to any specific model of M5-brane theory. 
As D4-branes are physically the same as M5-branes wrapping on a circle, 
we are tempted to assume it possible to 
``lift'' the gauge symmetry on D4-branes 
to the gauge symmetry on M5-branes. 
Since the Poisson limit 
(the lowest order approximation in the small $\theta$-expansion) 
of the NC gauge symmetry 
can be lifted to the NP gauge symmetry, 
it is natural to expect that the NC gauge symmetry 
can be lifted to a certain deformation of the NP gauge symmetry. 
However, 
we will prove that the answer to the question above is no. 
\footnote{
On the other hand, the NC D4-brane action can be directly 
obtained from the BLG model by choosing a Lie 3-algebra which 
does not resemble the NP algebra \cite{Np}.
}

\section{Review of NP Gauge Symmetry}
\label{review}

The Nambu-Poisson bracket is a natural generalization of the Poisson bracket. 
As the Poisson bracket 
\be
\{ f, g \} = \epsilon^{ij} \del_i f \del_j g
\qquad (i, j = 1, 2) 
\ee
can be used to generate the area-preserving diffeomorphisms in 2D by 
\be
\delta_A \phi = \{ A, \phi \}, 
\ee
the NP bracket 
\be
\{ A, B, C \} = \epsilon^{abc} \; \del_a A \; \del_b B \; \del_c C 
\qquad (a, b, c = 1, 2, 3)
\ee
can be used to generate 3D volume-preserving diffeomorphism (VPD) by 
\be
\delta_{(A, B)} \phi = \{ A, B, \phi \} 
\ee
for two arbitrary functions $A, B$. 
As we can linearly superpose transformations to obtain new transformations, 
a generic 3D transformation is of the form
\be
\delta_{\kappa} \phi = g \kappa^a \del_a \phi, 
\ee
where $\kappa^a$ are arbitrary functions satisfying the constraint
\be
\label{divless}
\del_a \kappa^a = 0. 
\ee
We can compute the Lie bracket of the 3D VPD 
\be
[ \delta_{\kappa}, \delta_{\kappa'} ] = \delta_{\kappa''},
\ee
where $\kappa''$ is found to be
\be
\kappa''{}^a = -g\left(
\kappa^b\del_b\kappa'{}^a
- \kappa'{}^b\del_b\kappa^a
\right),
\ee
which satisfies the condition (\ref{divless}) as required.

It was shown \cite{Ho:2008ve} that, via double dimension reduction,
the NP M5-brane action reduces to the Poisson limit of 
the NC super Yang-Mills action on 4+1 dimensions. 
Essentially this is because of the relation
\be
\{ A, B, x^3 \} = \{ A, B \}, 
\ee
where the left hand side is the NP bracket, 
and the right hand side is the Poisson bracket.
As the Poisson structure only accounts for the lowest order terms 
in the Moyal bracket 
defined for a noncommutative space
\begin{equation}
\label{MB}
[A(x),B(x)]_{\ast} = A\ast B -B \ast A
= i \theta \e^{ij}\del_iA \del_{j} B + {\cal O}(\theta^3),
\end{equation}
where the Moyal product $\ast$ is defined by
\be
A(x) \ast B(x) = \left. e^{\frac{i}{2}\theta\epsilon^{ij}\frac{\del}{\del x^i} \frac{\del}{\del y^j}} A(x) B(y) \right|_{y = x}
= A(x) B(x) + \frac{i}{2}\theta \{ A(x), B(x) \} + {\cal O}(\theta^2), 
\ee
we are looking for ways to recover the higher order terms in $\theta$ 
in order to obtain the full NC D4-brane action. 

There are several logical possibilities how a full NC D4-brane action can be obtained 
from a double dimension reduction of the M5-brane theory, 
depending on the extent to which a deformation is needed. 
The followings are the logical possibilities.
\begin{enumerate}
\item
\label{p1}
The NP M5-brane theory does not have to be deformed;
only the ansatz for double dimension reduction used in \cite{Ho:2008ve} 
should be deformed.
\item
\label{p2}
The NP bracket has to be deformed so that the NP gauge symmetry 
is deformed accordingly.
\item
\label{p3}
A more drastic deformation of the gauge symmetry is in need,
so that the notion of NP bracket has to be forsaken.
\end{enumerate}
Let us consider each possibility in more detail in turn.

\section{Deformation of Double Dimension Reduction}
\label{deformDDR}

The first possibility asks for the minimal deformation. 
Although it will be hard to convince everyone that one has exhausted 
all possible ansatz for double dimension reduction, 
it is actually possible to rule out the first possibility without any assumption about 
the double dimension reduction.
The reason is as follows.
Regardless of how double dimension reduction is implemented, 
if it is possible to obtain NC gauge symmetry 
from the gauge symmetry of 3D VPD 
via some sort of double dimension reduction, 
it must be possible to embed the NC gauge symmetry 
as a subgroup of the 3D VPD, 
because double dimension reduction is always
a restriction of the gauge symmetry to a subgroup. 
Let us now prove that it is indeed impossible 
to embed 2D NC gauge symmetry in 3D VPD.

Identifying NC gauge symmetry as a subgroup of 3D VPD means that 
there exists a map $\kappa({\lambda})$, 
which maps a NC gauge transformation parameter $\lambda$ to 
a VPD transformation parameter $\kappa$. 
We already know that the Poisson limit of the NC gauge symmetry is
a subgroup of the 3D VPD. 
The embedding map is
\be
\kappa_0^{\mu}(\lambda) = \left\{
\begin{array}{l}
\kappa_0^i(\lambda) = \epsilon^{ij}\del_j \lambda, \\
\kappa_0^3(\lambda) = 0,
\end{array}
\right.
\label{kappa0}
\ee
where $\mu = 1, 2, 3$, $i, j = 1, 2$, 
and we need to identify 
\be
\label{gtheta}
g = \theta.
\ee

To embed the NC gauge symmetry in 3D VPD, 
we need a deformation of the map above 
\be
\kappa(\lambda) = \kappa_0(\lambda) + \theta \kappa_1(\lambda) + \theta^2 \kappa_2(\lambda) + \cdots ,
\ee
such that for arbitrary $\lambda, \lambda'$,
\be
\kappa^{\mu}([\lambda, \lambda']_{\ast}) = - g\left(
\kappa^{\nu}(\lambda)\del_{\nu}\kappa^{\mu}(\lambda')
- \kappa^{\nu}(\lambda')\del_{\nu}\kappa^{\mu}(\lambda)
\right), 
\label{Closure sec 3}
\ee
where $g$ can be related to $\theta$ through a deformation of (\ref{gtheta})
\be
g = \theta + h_1 \theta^2 + h_2 \theta^3 + \cdots .
\ee

We will now prove that it is impossible to find $\kappa_1$ and $\kappa_2$ 
such that (\ref{Closure sec 3}) holds for all $\lambda, \lambda'$. 

The NC gauge symmetry algebra is defined by the Moyal bracket (\ref{MB}), 
and can be expressed as a $\theta$-expansion
\be
[\lambda, \lambda']_{\ast} 
= \theta \{\lambda, \lambda'\} + \theta^2 M_1(\lambda, \lambda') + \theta^3 M_2(\lambda, \lambda') + \cdots ,
\ee
where $M_1(\lambda, \lambda') = 0$ and 
\be
M_2(\lambda, \lambda') = \frac{1}{3}\epsilon^{ij}\epsilon^{kl}\epsilon^{mn}
(\del_i\del_k\del_m \lambda) (\del_j\del_l\del_n \lambda') .
\ee

In general, the maps $\kappa_n(\lambda)$ is linear in $\lambda$, 
and in principle they can depend on the gauge potential or other fields of the gauge theory. 
However, the closure relation (\ref{Closure sec 3}) should be satisfied 
regardless of the values of these fields, 
thus we can set all other fields to zero and consider $\kappa_n$'s as 
pseudodifferential operators on $\lambda$ as
\be
\kappa_n^{\mu}(\lambda) = \sum_{k = 0}^{\infty} 
K_{nk}^{\mu |\mu_1\cdots \mu_k}(x)
(\del_{\mu_1}\cdots\del_{\mu_k} \lambda). 
\ee

Let us now try to solve $\kappa_n$ order by order from (\ref{Closure sec 3}). 
The lowest order terms are 
\be
\kappa_0^{\mu}(\{\lambda, \lambda'\}) = - \left(
\kappa_0^{\nu}(\lambda)\del_{\nu}\kappa_0^{\mu}(\lambda')
- \kappa_0^{\nu}(\lambda')\del_{\nu}\kappa_0^{\mu}(\lambda)
\right).
\ee
This identity is already satisfied by our choice of $\kappa_0$ (\ref{kappa0}). 

At the next order, (\ref{Closure sec 3}) gives 
\be
\kappa_1^{\mu}(\{\lambda, \lambda'\}) = h_1 \kappa_0^{\mu}(\{\lambda, \lambda'\})
- \left(
\kappa_0^{\nu}(\lambda)\del_{\nu}\kappa_1^{\mu}(\lambda')
+ \kappa_1^{\nu}(\lambda)\del_{\nu}\kappa_0^{\mu}(\lambda')
- (\lambda \leftrightarrow \lambda')
\right).
\label{kappa1eq}
\ee
Expanding $\kappa_1$ as 
\be
\kappa_1^{\mu}(\lambda) = 
\sum_{k=0}^{\infty} F_k^{\mu|\mu_1\cdots \mu_k}\del_{\mu_1\cdots \mu_k}\lambda,
\ee
we notice that only $F_3$, and $F_4$ will be relevant 
for the argument below. 

A necessary condition for (\ref{kappa1eq}) to hold is that 
all the terms of the form $\del^3\lambda\del^2\lambda'$ cancel.
Restricting $\lambda$ and $\lambda'$ to be independent of $x^3$, 
we then find
\be
3F_3^{i|jkl}\{\del_j \del_k \lambda, \del_l\lambda'\} = 
- \epsilon^{in}F_3^{j|klm}\del_k\del_l\del_m \lambda \del_j\del_n\lambda'.
\ee
For $\lambda = e^{ik\cdot x}$ and $\lambda' = e^{ik'\cdot x}$, 
this implies that 
\be
k'_i F_3^{i|jkl} k_j k_k k'_l = 0, 
\ee
which is solved by
\be
F_3^{i|jkl}k_j k_k = \alpha\epsilon^{il}
\label{anti-symmetric}
\ee
where $\alpha$ can be an arbitrary function.
However, by definition 
$F_3^{i|jkl}$ must be symmetric with respect to $(j, k, l)$, 
and since there is no way to preserve the relation above 
while symmetrizing $j, k, l$, 
the only possibility is that $F_3^{i|jkl} = 0$.
\footnote{
If $F_{3}^{i|jkl}$ and $F_{4}^{i|jklm}$ obeys 
(\ref{anti-symmetric}) and (\ref{F4-2}) respectively,
$F_3^{i|jkl}$ and $F_4^{i|jklm}$ must both vanish 
because
\begin{equation}
F_{3}^{i|jkl} = -F_{3}^{l|jki} = 
-F_{3}^{l|jik} = F_{3}^{k|jil} = F_{3}^{k|jli} 
= -F_{3}^{i|jlk} = -F_{3}^{i|jkl},
\end{equation}
\begin{equation}
F_{4}^{i|jklm} = -F_{4}^{m|jkli} = 
-F_{4}^{m|jkil} = F_{4}^{l|jkim} = F_{4}^{l|jkmi} 
= -F_{4}^{i|jkml} = -F_{4}^{i|jklm}.
\end{equation}
}

Similarly, for $F_4$, we consider terms of the form $\del^4\lambda\del^2\lambda'$, 
\bea
4 F_4^{i|jklm}\{\del_j \del_k \del_l \lambda, \del_m\lambda'\} &=& 
- \epsilon^{ir}F_4^{j|klmn}\del_k\del_l\del_m\del_n \lambda \del_j\del_r\lambda'.
\eea
For $\lambda = e^{ik\cdot x}$ and $\lambda' = e^{ik'\cdot x}$, 
this implies
\begin{equation}
k_i'F_4^{i|jklm} k_jk_kk_lk_m' = 0
\end{equation}
which is solved by
\begin{equation}
F_4^{i|jklm} k_jk_kk_l = \beta \e^{im}, \label{F4-2}
\end{equation}
where $\beta$ is just a constant.
Since $F_4^{i|jklm}$ is totally symmetrized among $jklm$ by definition, 
(\ref{F4-2}) requires that $F_4^{i|jklm} = 0$ identically.

At order $\theta^4$, (\ref{Closure sec 3}) gives 
\bea
\kappa_2^{\mu}(\{\lambda, \lambda'\}) 
&=&
- \kappa_0^{\mu}(M_2(\lambda, \lambda')) 
+ h_2 \kappa_0^{\mu}(\{\lambda, \lambda'\})
- h_1\left(
\kappa_0^{\nu}(\lambda)\del_{\nu}\kappa_1^{\mu}(\lambda')
+ \kappa_1^{\nu}(\lambda)\del_{\nu}\kappa_0^{\mu}(\lambda')
- (\lambda \leftrightarrow \lambda')
\right) 
\nn \\
&&
- \left(
\kappa_1^{\nu}(\lambda)\del_{\nu}\kappa_1^{\mu}(\lambda')
+ \kappa_0^{\nu}(\lambda)\del_{\nu}\kappa_2^{\mu}(\lambda')
+ \kappa_2^{\nu}(\lambda)\del_{\nu}\kappa_0^{\mu}(\lambda')
- (\lambda \leftrightarrow \lambda')
\right).
\label{kappa2eq}
\eea
Let us now check whether non-trivial $\kappa_2$ exists. 
Consider the case $\mu = i$. 
The first term on the right hand side is of the form $\del(\del^3\lambda\del^3\lambda')$.
This term has no match on the right hand side, 
as each term on the right hand side has less than 3 derivatives on either $\lambda$ or $\lambda'$.
A priori such a term could arise from the term 
$\kappa_1^{\nu}(\lambda)\del_{\nu}\kappa_1^{\mu}(\lambda')$, 
but this can not happen as we have proven above that $F_3^{i|jkl} 
= F_4^{i|jklm} = 0$.
As a result, 
the first term on the right hand side has to be cancelled by the term
on the left hand side.
For the special case 
\be
\lambda = e^{ik\cdot x}, \qquad \lambda' = e^{ik'\cdot x}, 
\ee
the 1st term on the right hand side and the term on the left hand side 
of (\ref{kappa2eq}) are 
\be
-i\frac{g^2}{3}\epsilon^{ij}(k+k')_j(k\times k')^3 e^{i(k+k')\cdot x}
\qquad \mbox{vs} \qquad 
g(k\times k')\kappa_2^i(e^{i(k+k')\cdot x}).
\ee
Apparently $\kappa_2$ has to be a pseudo-differential operator, 
and 
$\kappa_2^i(e^{i(k+k')\cdot x}) = f^i(k+k')e^{i(k+k')\cdot x}$ 
for some function $f^i$. 
After a change of variables 
\be
p = k+k', \qquad q = (k'-k)/2,
\ee 
the two terms become
\be
-i\frac{g^2}{3}\epsilon^{ij}p_j(p\times q)^3 e^{ip\cdot x} \qquad 
\mbox{vs} \qquad 
g(p\times q) f^i(p) e^{ip\cdot x},
\ee
hence we need 
\be
f^i(p) = i\frac{g}{3} \epsilon^{ij} p_j (p\times q)^2 
\ee
for (\ref{kappa2eq}) to hold.
Yet this is impossible because $f^i(p)$ is independent of $q$.
Therefore $\kappa_{2}$ can not exist.
The conclusion is that NC gauge symmetry can not be 
a subgroup of 3D VPD.

\section{Deformation of the NP Bracket}

The next simplest possibility of obtaining a new theory for finite $C$-field background 
is that there exists a deformation of the Nambu-Poisson bracket 
analogous to the Moyal bracket, 
which can be viewed as a deformation of the Poisson bracket. 
In order for the deformed Nambu-Poisson bracket to be ready to define a gauge symmetry, 
the new bracket must satisfy the fundamental identity
\be
{}[A, B, [C, D, E]] = [[A, B, C], D, E] + [C, [A, B, D], E] + [C, D, [A, B, E]].
\ee
In terms of the definition of a transformation generated by the deformed NP-bracket
\be
\label{transf}
\delta_{(A, B)} = [A, B, \cdot], 
\ee
the fundamental identity can be understood as the covariance of the bracket
\be
\delta_{(A, B)}[C, D, E] = [\delta_{(A, B)}C, D, E] + [C, \delta_{(A, B)}D, E] + [C, D, \delta_{(A, B)}E]. 
\ee
It can also be rewritten as the closure of the algebra of transformations as
\be
[\delta_{(A, B)}, \delta_{(C, D)}]E 
= \delta_{([A, B, C], D)+(C, [A, B, D])} E.
\label{NPclosure}
\ee
This identity is usually referred to as the fundamental identity. 
If the 3-bracket $[A, B, C]$ is assumed to be totally antisymmetric, 
the algebra satisfying (\ref{NPclosure}) is called a Lie 3-algebra, 
and it can be used for the ${\cal N} = 8$ BLG model. 
If the 3-bracket $[A, B, C]$ is antisymmetric only in $A, B$, 
the algebra satisfying (\ref{NPclosure}) can be used to define 
an ${\cal N} = 6$ model in 3D \cite{N=6}. 
In the literature \cite{tak,Dito}, a deformation of the 
NP bracket often assumes 
the Leibniz rule
\be
[ A, B, CD ] = [ A, B, C ] D + C [ A, B, D ].
\ee

But it turns out that
the 2nd possibility, 
which is the theme of this section, 
can be ruled out even without assuming the Leibniz rule.
Actually, we will see in the next section that, 
without any assumption about the form of the gauge symmetry 
except that it is a deformation of the 3D VPD 
(the 3rd possibility discussed at the end of sec. \ref{review}),
it is impossible to obtain NC gauge symmetry 
upon double dimension reduction of any sort. 
Hence the possibility that a deformed NP bracket 
reduces to the Moyal bracket is also ruled out.

\section{Deformation of the Gauge Symmetry}

In general, it is interesting to know whether 
the algebra of $d$ dimensional VPD admits any nontrivial deformations. 
As any algebra of symmetry with infinitesimal transformation laws, 
the deformed symmetry must have the structure of a Lie algebra. 
In this case, we are looking for an infinite dimensional Lie algebra 
satisfying the Jacobi identity 
\be
[[X, Y], Z] + \mbox{cyclic} \equiv [[X, Y], Z] + [[Y, Z], X] + [[Z, X], Y] = 0, 
\ee
as a deformation of the $3$ dimensional VPD.

Assuming that there is a deformation of the Lie bracket $[\cdot, \cdot]_0$ 
of a Lie algebra ${\cal G}$, 
one can expand the deformed Lie bracket as 
\be
[X, Y]_g = [X, Y]_0 + g c_1(X, Y) + g^2 c_2(X, Y) + \cdots,
\ee
where $g$ is the deformation parameter 
and $c_1, c_2, \cdots$ are bilinear functions on ${\cal G}$.
The Jacobi identity has to be satisfied order by order. 
While it is trivial at the 0-th order, 
the Jacobi identity at the 1st order is
\be
\delta c_1(X, Y, Z) \equiv [c_1(X, Y), Z]_0 
+ c_1([X, Y]_0, Z) + \mbox{cyclic} = 0.
\ee

Note that the 1st order deformation would be considered trivial if 
it merely corresponds to a change of basis of the Lie algebra generators. 
That is, a linear transformation on ${\cal G}$
\be
X \rightarrow \Phi(X) = X + g b(X) + {\cal O}(g^2), 
\ee
where $b$ is a linear map on ${\cal G}$, 
is not considered to change the Lie algebra, 
although it induces a change of the appearance of the Lie bracket 
\be
[\Phi(X), \Phi(Y)]_0 = \Phi([X, Y]_0) + g \delta b(X, Y) + {\cal O}(g^2), 
\ee
where 
\be
\delta b(X, Y) \equiv [b(X), Y]_0 + [X, b(Y)]_0 - b([X, Y]_0).
\ee
Therefore, if $c_1 = \delta b$ for some $b$, 
$c_1$ is considered a trivial deformation. 
Furthermore, if two choices of $c_1$ are considered equivalent 
if their difference is $\delta b$ for some $b$. 
In short, $c_1$ should be considered as an element 
in the cohomology defined by $\delta$, 
denoted by $H^2({\cal G}, {\cal G})$. 
A necessary condition for the existence of nontrivial deformations of ${\cal G}$ 
is then the cohomology class $H^2({\cal G}, {\cal G})$ is nontrivial. 

In \cite{rigidity}, it was shown that $H^2({\cal G}, {\cal G})$ is trivial 
for ${\cal G}$ being the VPD on compact manifolds in 4 or higher dimensions. 
For VPD on 3 dimensional compact manifolds, 
$H^2({\cal G}, {\cal G})$ is 1 dimensional. 
However, even if $H^2({\cal G}, {\cal G})$ is nontrivial, 
it does not imply the existence of nontrivial deformation of ${\cal G}$. 
One still needs to check that the Jacobi identity holds at higher orders. 
At the 2nd order, 
the Jacobi identity implies that 
\be
- \left[ c_1(c_1(X, Y), Z) + \mbox{cyclic} \right] = \delta c_2(X, Y, Z). 
\ee
Thus a nontrivial deformation of the 3D VPD 
exists only if the left hand side of the equation above is
a trivial element in $H^3({\cal G}, {\cal G})$ for 
the nontrivial element $c_1$ (the only one) in $H^2({\cal G}, {\cal G})$.
This turns out to be not true, 
and the conclusion is that the VPD is {\em rigid} for all $d \geq 3$, 
i.e., it does not admit any nontrivial deformation \cite{rigidity}. 
\footnote{
2D VPD admits deformation, e.g.
the noncommutative gauge symmetry generated via the Moyal bracket.
}

The rigidity of the gauge symmetry of 3D VPD 
implies that there is no nontrivial deformation of the Nambu-Poisson bracket. 

Strictly speaking, this theorem only applies to compact manifolds,
\footnote{
In \cite{Dito} a quantum version of the Nambu-Poisson bracket is 
constructed based on the so-called Zariski algebra defined for $\mathbb{R}^3$. 
It remains to be checked whether this algebra leads to a deformation of VPD. 
If it does, it provides an example of deformed 3D VPD for noncompact space. 
}
that is, to the case when the M5-brane worldvolume is a product space 
${\cal M}_{1+2} \times {\cal N}_3$ 
where ${\cal N}_3$, the 3D space on which the Nambu-Poisson bracket is defined, 
is compact. 
It is possible that a noncompact ${\cal N}_3$ can still admit 
a deformation of its VPD. 
However, in a physical theory, finite-energy physical states are local configurations 
which decay to zero at infinity. 
In this sense, the noncompact space $\mathbb{R}^3$ should be treated 
the same as the infinite-volume limit of, say, $S^3$ or $T^3$. 
Thus we expect that the no-go theorem also applies to M5-branes 
with non-compact base space, 
at least when certain locality conditions are imposed.


\section{Conclusion}

We have proven in the above the no-go theorem that 
it is impossible to deform the NP gauge symmetry so that 
it includes the NC gauge symmetry as a subgroup to all orders. 
The basic reason of this no-go theorem is that 
the NP gauge symmetry (for compact ${\cal N}_3$)
is precisely the 3 dimensional volume-preserving-diffeomorphism (VPD). 
While the 3D VPD does not include the NC gauge symmetry in 2D as a subgroup. 
It is also rigid against deformation. 

The rigidity of the 3D VPD implies that the NP M5-brane theory is also rigid, 
in the sense that it can only be deformed by adding gauge invariant terms 
to its Lagrangian without deforming its gauge transformation laws. 
This means that we can not directly identify the gauge symmetry of an NC D4-brane 
with the NP gauge symmetry of an M5-brane, 
regardless of how we deform the NP M5-brane theory. 

On the other hand, the no-go theorem does not imply that it is 
impossible to construct an M5-brane theory that reduces to 
the D4-brane theory in $B$-field background.
In order to find the correction terms in the NP M5-brane theory 
to reproduce NC D4-brane theory, 
one can first use a Seiberg-Witten map to rewrite the NC D4-brane action 
in terms of the gauge fields with Poisson gauge symmetry, 
and then the gauge transformation laws of the D4 and M5-brane theories 
can be directly identified through the double dimension reduction used in \cite{Ho:2008ve}.
The correction terms will be easy to find, 
at least order by order, after the gauge symmetries are matched.

The moral of the no-go theorem presented in this paper is that there is still a lot
to be learned about the M5-brane physics, 
even before we take up the challenge of the problem of multiple M5-branes.

\section*{Acknowledgements}

The authors thank Wei-Ming Chen, Chong-Sun Chu, Kazuyuki Furuuchi,
Andreas Gustavsson,
Hiroshi Isono, Yutaka Matsuo, Darren Sheng-Yu Shih, Dan Tomino,
Chi-Hsien Yeh for helpful discussion.
T.T would like to thank Yukawa Institute for Theoretical Physics,
in particular Prof.~Hiroshi Kunitomo, for the hospitality and support
during his stay in October 2009.
This work is supported in part by
the National Science Council,
and the National Center for Theoretical Sciences, Taiwan, R.O.C.

\providecommand{\href}[2]{#2}\begingroup\raggedright\endgroup

\end{document}